\documentclass[notitlepage,prd,onecolumn,showpacs,preprintnumbers,floatfix,nofootinbib,superscriptaddress]{revtex4-2}
\usepackage{amssymb,amsmath,placeins,epsfig,array}
\usepackage[usenames,dvipsnames]{color}
\usepackage[normalem]{ulem}
\usepackage[breaklinks,colorlinks,urlcolor=blue,citecolor=blue,linkcolor=blue]{hyperref}
\usepackage{lipsum}
\usepackage{lineno}
\usepackage{float}
\usepackage{mathptmx}
\include{commands}

\begin{document}

\title{Local Lorentz invariance tests for photons and hadrons \\ at the Gamma Factory}

\author{B.~Wojtsekhowski} 
\affiliation{William \& Mary, Williamsburg, Virginia, 23187, USA \\
Thomas Jefferson National Accelerator Facility{,} Newport News{,} Virginia 23606, USA\\} 
\author{Dmitry Budker}
\affiliation{Johannes Gutenberg-Universit{\"a}t Mainz, 55128 Mainz, Germany\\
Helmholtz-Institut, GSI Helmholtzzentrum f{\"u}r Schwerionenforschung, 55128 Mainz, Germany\\
Department of Physics, University of California, Berkeley, California 94720, USA}

\begin{abstract}
High-precision tests of local Lorentz invariance, via monitoring of the sidereal time variation of the photon energies emitted by ultrarelativistic heavy-ion beams and of the beam momentum, are proposed. 
This paper includes descriptions of the physics ideas and the concept for the detector.
The experiment results will allow high-precision tests of LLI via anisotropy of the maximum attainable speed of a photon and an ion. 
The projected accuracy for the asymmetries interpreted in the framework of the anisotropic relativistic mechanics corresponds to the limit on sidereal time variation of the one-way maximum attainable speed at the levels between $10^{-14}$ and $10^{-17}$.
\end{abstract}
\maketitle

\section{\label{sec:Introduction}Introduction}

Isotropy of the speed-of-light, a key postulate of the special theory of relativity (STR), has been tested in many experiments over almost 150 years, the precision improving by many orders, see reviews~\cite{Mattinly2005ab, AKostel2011ab, Will2014ab}.
These experiments belong to a larger group of investigations related to tests of local Lorentz invariance (LLI).
Advanced theoretical models have been developed~\cite{HRobert1949ab, Mansouri1977ab, Colladay1997a} for accurate interpretation 
of the experimental limits on LLI violation.
Many experiments and observations have been analyzed in the framework of the Standard Model extension (SME), see Ref.\,\cite{AKostel2011ab}.
There are two classes of experiments on speed-of-light anisotropy. 
Experiments in the first class, starting from the famous work of Ref.\,\cite{MM1887ab}, test the round trip (two-way) speed, $c2$.
The second class of experiments addresses the one-way speed-of-light, $c1$.
In STR, the speed-of-light is the maximum attainable speed for all types of matter. 
It was suggested about 20 years ago~\cite{Colladay1998b, Col1999ab} that particles could have different values of the maximum attainable speed,
so many additional experiments would be useful.

In this paper, we present three schemes for testing LLI.
Two tests are based on the correlations between the energies of two photon beams moving in the same or opposite directions and the third test is based on a correlation between the photon energy and the ion beam momentum.

\section{\label{sec:Fast}Special relativity tests with beams of fast particles}

The relativistic Doppler effect (DE) was already considered by A.~Einstein in his 1905 paper \cite{AE1905ab}.
Doppler effect provided a way to observe a key feature of relativistic space-time: the time-dilation effect.
The longitudinal DE, originating in time-dilation, results in a change in photon frequency due to Lorentz boost.
The first such test was successfully realized by Ives\&Stilwell~\cite{Ives1938ab} using a beam of fast atoms.
With progress in photon and accelerator technology, a number of experiments have been performed, see reviews~\cite{Mattinly2005ab, Will2014ab}.

Thanks to modern experimental methods, 
the most advanced experiment using ions in a storage ring was completed in 2014~\cite{Botermann2014ab}, which directly tested the DE formula and
achieved sensitivity to a speed--of--light variation at the level of $10^{-8}$.
The anisotropy of maximum attainable speed (MAS) via the sidereal time variation of DE could also be tested, but this requires a very stable energy of the ion beam. 
Hereafter, we referring to the average value of the energy of the ion beam because the ion beam energy spread is small enough for considerations presented below.

Photon scattering from a fast moving electron (inverse Compton scattering) has also been proposed to search for sidereal time variation 
of the photon MAS~\cite{Gurzadyan1996ab}.
The large value of Lorentz factor of the electron beam ($\gamma$) made possible a high-precision test due to the enhanced sensitivity of the final
photon energy ($E$) to the variation of the speed-of-light: $\delta E/E = 2 \, \gamma^2 (\delta c/c)$.
The result of the experiment~\cite{Bocquet:2010ke}, interpreted in the framework of SME, provided a limit on the anisotropy of the one-way MAS at the level of $10^{-14}$.

High-energy particle beams in a storage ring can be used for precision tests of the Lorentz-force variation with sidereal time, 
especially with two beams moving in opposite directions in the same magnetic structure \cite{Wojt2014ab, Bergan:2019lff}.
In such an experiment the use of two counter-propagating beams provides a way to reduce the impact of the magnetic system instability 
and beam-energy variations.
The sensitivity to a potential variation of the electron MAS is enhanced by an additional factor of 2. 
The result of that experiment interpreted in the framework of anisotropic relativistic mechanics (ARM)~\cite{Sonego2009ab} as a limit on the anisotropy of the electron MAS (in this case, $c1_e$)
is $5 \cdot 10^{-15}$.

\section{Accelerator experiment with partially-stripped ions}

\begin{figure}[htb]
\centering
\includegraphics[trim = 0mm 110mm 0mm 30mm, width=0.95\textwidth]{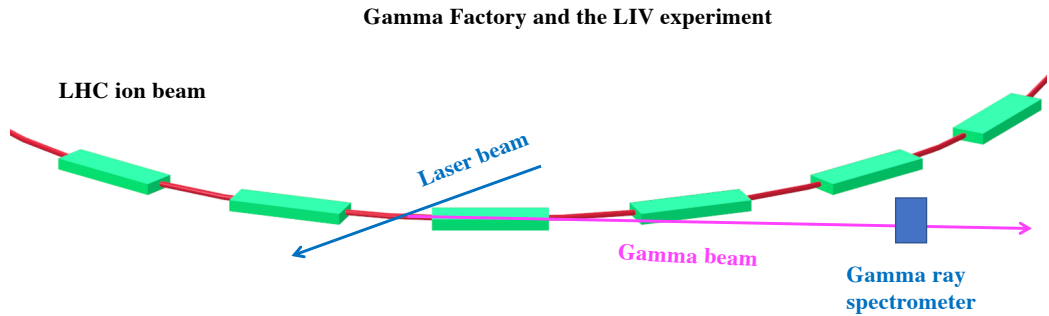} 
\caption{Concept view of the experiment at the Gamma Factory.}
\label{fig:layout}
\end{figure}

Here we present a new possibility for a DE experiment using a photon beam resulting from resonant light scattering by high-energy partially-stripped ions; for the anticipated photon-beam parameters, see Refs.~\cite{krasny2015gamma,Krasny2018,ISPIRIAN1973377,budker2020atomic}. 
A hydrogenic ion beam in LHC could allow a total photon intensity of $10^{17}$ per second, thanks to the large cross section of the photon-atom interaction on resonance, and presents a unique opportunity for an advanced DE experiment, shown schematically in Figure\,\ref{fig:layout}.

The Lorentz factor ($\gamma_{beam}$) for heavy ions achievable at LHC is up to 2900.
The process which leads to high-energy photons in the proposed Gamma Factory~\cite{Krasny2018} relies on the relativistic DE for the excitation of, for example, hydrogen-like ions on, for example, 1s$\rightarrow$2p and emission on 2p$\rightarrow$1s as, shown in Figure\,\ref{fig:atom}.

\section{Two-step de-excitation of the partially-stripped ions}
\label{sec:configuration_A}
\begin{figure}[htb]
\centering
\includegraphics[trim= 0mm 0mm 0mm 0mm, width=0.45\textwidth]{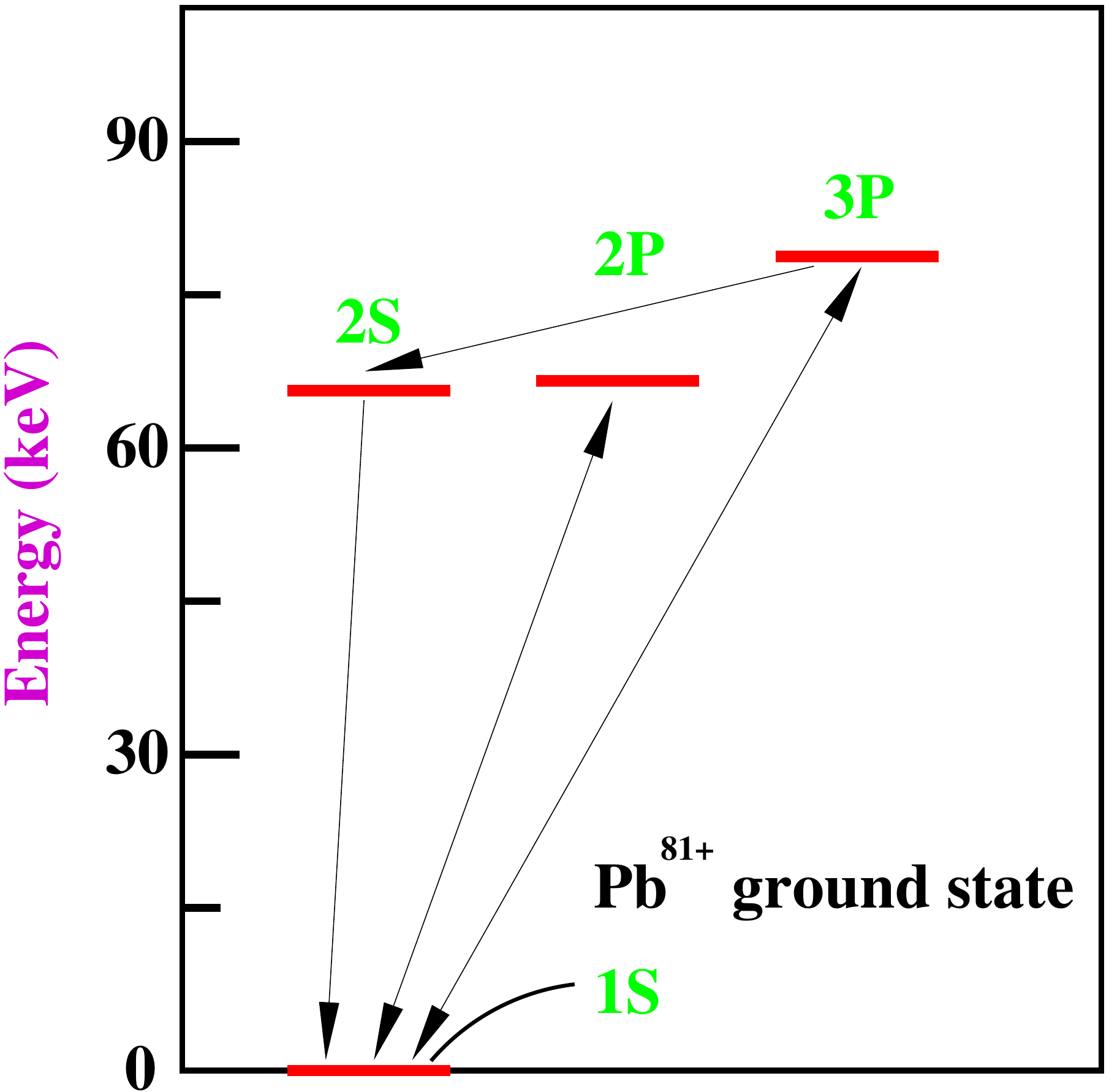}
\caption{Energy-level scheme of a hydrogen-like ion (exemplified by lead) and transitions in the proposed experiment.}
\label{fig:atom}
\end{figure}

At the GF, one could also use a different scheme of atomic transitions:
1s$\rightarrow$3p with the following de-excitation via the  3p$\rightarrow$1s transition
or via two transitions: 3p$\rightarrow$2s and 2s$\rightarrow$1s, see Figure\,\ref{fig:atom}.

The advantage of the latter option is the different dependence of the photon energies for the 3p1s and 2s1s cases on the ion-beam energy. 
The energies ($E_{131}$ and $E_{1321}$) of the two secondary photon lines are predicted by STR:
\begin{equation}
E_{1321} = E_{2s} \times \sqrt {E_{131}/E_l}, 
\end{equation}
where $E_{2s}$ is the energy of the 2s$\rightarrow$1s transition  and $E_l$ is the energy of the laser photon. Correlation between these energies could be used for a test of the LLI.

In the case of the 1s$\rightarrow$3p$\rightarrow$1s process, there are two Lorentz boosts.
The first boost is from the lab reference frame to the ion-beam reference frame, and the second boost is from the ion-beam reference frame to the lab reference frame.
The Lorentz factor is defined by the difference between the speed of the ion beam (the ion-beam reference frame) relative to the lab reference frame and the speed of the photon in the direction of its motion (here we are using a formulation similar to one of the Ref.\,\cite{Sonego2009ab}, see section 2.1.1).
Because in the process 1s$\rightarrow$3p$\rightarrow$1s the photon changes its direction of motion, the resulting photon energy, defined by the product of two boosts, is sensitive to the value of the two-way speed-of-light.
\newpage
In the case of the 1s$\rightarrow$3p$\rightarrow$2s$\rightarrow$1s process,
the photon from the final transition (2s$\rightarrow$1s) is subject to only one Lorentz boost, so its energy is sensitive to the value of the one-way speed-of-light.

\subsection{Rates of atomic transitions}
The photon energy required to excite a $Z=81$ hydrogenic ion is about 78\,keV in the rest frame of the ion or 13.4\,eV in the laboratory frame.
Below, we are using calculations per Refs.\,\cite{PhysRevA.77.042507, SURZHYKOV2005139,  MCCONNELL2010711, popov2017relativistic} for atomic transition rates.
The transition probability for the 2p$_{1/2} \rightarrow$1s transition is $3.0 \cdot 10^{16}$ per second.
We will use excitation of the 3p$_{1/2}$ state.
The transition probability for 3p$_{1/2} \rightarrow$1s is $6.9 \cdot 10^{15}$ per second.
The two-step de-excitation proceeds via the 3p$_{1/2}\rightarrow$2s transition with rate $1.3 \cdot 10^{15}$ per second
and the  2s$\rightarrow$1s transition with rate $5.3 \cdot 10^{13}$ per second.
The branching fraction for the two-step de-excitation is about 16\%.
An important feature of high-Z hydrogenic ions is the dominance of one-photon (M1) transition in de-excitation of the 2s atomic state~\cite{andreysur}.
Taking into account the time dilation factor, we find that the lifetime of the excited 2s state is $0.5 \cdot 10^{-10}$\,s in the laboratory frame.
These transition probabilities are sufficiently high that the excitation and de-excitation of the ion will happen within a few cm of the point of interaction between the ions and the laser light. 

\subsection{Beam-energy variation and its correction}
\label{sec:energy}

An issue that needs to be resolved for a sensitive LLI study at the GF is the significant daily variation of the beam energy known for the LHC tunnel
since the Z-boson LEP era~\cite{ARNAUDON1995249}.
The amplitude of this variation ($\Delta \gamma_{beam}/\gamma_{beam}^{average}$) is on the scale of $10^{-4}$ and not exactly predictable due to contributions from several tidal effects and weather conditions.

The energy of the final photon ($E_{_{131}}$) (after 1s$\rightarrow$3p$\rightarrow$1s transitions) can be presented as:
\begin{equation}
E_{_{131}} =  4\gamma_{beam}^2 \cdot E_l = E_{_{131}}^{average} \left ( 1 + 2 \frac{\Delta \gamma_{beam}}{\gamma_{beam}^{average}} \right ),
\end{equation}
where $E_l$ is the energy of the primary laser photon.
We took into account that for the selected atomic transition the width of the excited state is larger than photon energy spread in the ion rest frame.
When the beam energy varies, the energy of the emitted photon also changes, which makes it difficult to search for a tiny sidereal variation
related to the violation of LLI in spite of the $\gamma^2$ enhancement of sensitivity to the variation of the photon MAS.
In framework of ARM the $E_{_{131}}$ is sensitive to the two-way MAS of photon
(denoted here as $c2_{ph}$):
\begin{equation}
E_{_{131}} =  E_{_{131}}^{average} \left ( 1 + 2 \frac{\Delta \gamma_{beam}}{\gamma_{beam}^{average}} + 2\, \gamma^2 \frac{\delta c2_{ph}}{c2_{ph}} \right ) .
\label{eq:131}
\end{equation}
Due to the very tight experimental limit on the $\delta c2_{ph}/c2_{ph} \leq 10^{-18}$ reported in Ref.~\cite{Nagel_2015}, the last term in Equation~\ref{eq:131} could be ignored for the current analysis.

We propose to use two distinct final photon lines in the DE experiment and evaluate (again in framework of ARM) the combination of the corresponding photon energies in a search for a sidereal time variation of the photon energy~\cite{BW2020ab}.
As shown above, the energy for the final photon originating from the 1s$\rightarrow$3p$\rightarrow$1s process is proportional to the beam Lorentz factor squared.
At the same time, the energy for the final photon from 1s$\rightarrow$3p$\rightarrow$2s$\rightarrow$1s is proportional to only the first power of the Lorentz factor 

\begin{equation}
E_{_{1321}} =  2\gamma_{beam} \cdot E_{2s1s} = E_{_{1321}}^{average} \left ( 1 + \frac {\Delta \gamma_{beam}}{\gamma_{beam}^{average}} \right ) .
\end{equation}
Effectively, the variation of the energy of the incident photon in the ion rest frame does not affect the energy of the 2s atomic state.

The term related to the MAS of a photon for the two-step de-excitation (1321 case) is included now:
\begin{equation}
E_{_{1321}} =  E_{_{1321}}^{average} \left (1 + \frac{\Delta \gamma_{beam}}{\gamma_{beam}^{average}} + \gamma^2 \frac{\delta c1_{ph}}{c1_{ph}} \right ) .
\label{eq:1321}
\end{equation}
Because the energy of the second emitted photon in the  1s$\rightarrow$3p$\rightarrow$2s$\rightarrow$1s process is related to a one Lorentz boost, the study of the sidereal time variation of that energy allows one to access the anisotropy of the one-way maximum attainable speed of a photon.
The following formula can be used for the search of the photon one-way MAS variation: 
\begin{equation}
 \frac{\delta c1_{ph}}{c1_{ph}} = \gamma^{-2} \times F_{E-corr},
 \label{eq:c1}
\end{equation}
where $F_{E-corr}$ is the factor corrected for the beam energy variation defined as:
\begin{equation}
F_{E-corr} =  \left \{ \left[ \frac{E_{_{1321}}}{E_{_{1321}}^{average}} -1 \right] - 0.5 \left [ \frac{E_{_{131}}}{E_{_{131}}^{average}} -1 \right ] \right \} .
\label{eq:energy}
\end{equation}

In the framework of SME, use of two photon energies also allows the search for a variation of MAS of light because both energies have the same term $2 \gamma^2 \frac{\delta c1_{ph}}{c1_{ph}}$ 
and the beam energy variation effect could be found again from Equation~\ref{eq:energy}.

\subsection{Photon detector}

In this experiment, thanks to the high intensity of the photon beam we can use a detector with low efficiency but high energy resolution.
A specific detector will be selected based on the photon energy.
For the case of Z=81 the photon energy ($E_\gamma$) is high and we plan to realize
a magnetic 180 degree pair spectrometer, see, for example, Ref.~\cite{Golubnichiy:1969ui}.
The photon-energy resolution of such spectrometer used in previous work was 1\% limited by scattering of the pair components in the converter. 
However, for the proposed experiment, the converter thickness will be $10^{-6}$ of the radiation length (0.5\,$\mu$m of carbon) and the angular spread of the pair components will be defined by production process on the level of $m_e/E_\gamma$, where $m_e$ is the electron mass.

The coordinate detector will be based on Si microstrips with coordinate resolution of 10\,$\mu$m and time resolution of 2.5\,ns, see, for example, Ref.\,\cite{BATTAGLIERI201591}.
These allow, with a one-meter-size spectrometer, to reach an energy resolution of $10^{-4}$ (consistent with the contribution from the ion beam energy spread and the photon beam collimation) and a rate capability of 10\,MHz.
In addition to high energy resolution, the pair spectrometer has a low $\gamma \rightarrow e^+e^-$ pair conversion 
efficiency, which allows one to measure the energy spectrum for individual photons, even if they are emitted as a pair.
Figure~\ref{fig:PS} shows the cutaway view of the detector system with vertical orientation of the magnet gap.

The trigger will use the signals from the calorimeters.
The stream type readout system will use the coordinates of the hits in the microstrips and provides a sum of the hit's coordinates which equals to the photon energy. 

The experiment can use lower photon energy, for example, in the few-MeV energy range where even more precise technique for measurement of the photon energy was developed~\cite{PhysRevC.73.044303}.

\subsection{Data collection and projected results}

The expected total photon rate at the GF is up to $10^{17}$ photons per second for the 131 line and 10 times less for the 1321 branch.
\begin{figure}[htb]
\centering
\includegraphics[trim= 0mm 0mm 15mm 0mm, width=0.38\textwidth]{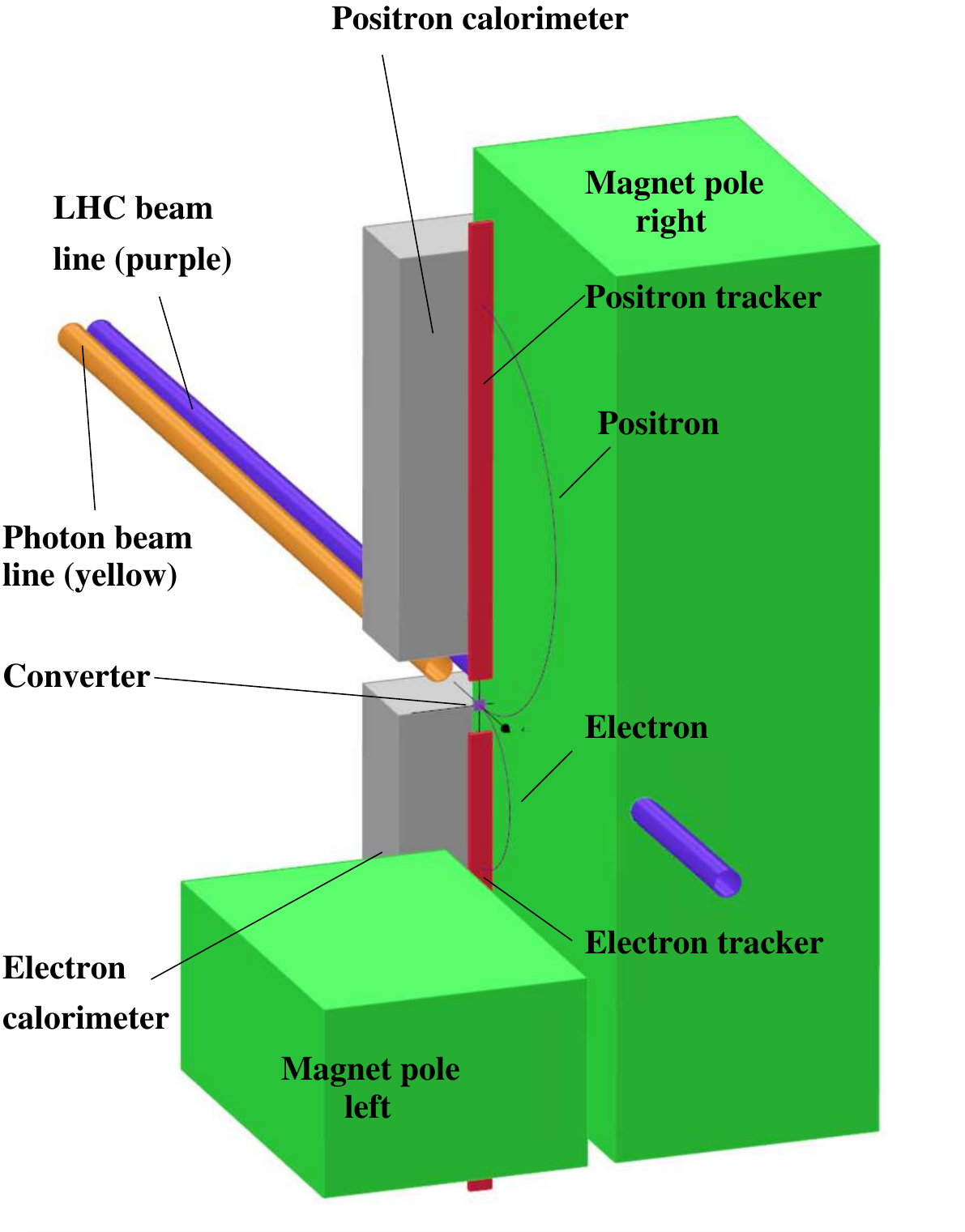}
\caption{A cut-off view of the pair spectrometer with vertical magnetic gap.}
\label{fig:PS}
\end{figure}
Such rate is too high for any detector even with an efficiency of $10^{-6}$, so we will need to reduce it by collimation of the photon beam, which is also required for the selection of a narrow range of the photon energy.
Alternatively, the converter can be made in the form of a disk, so the photons with large angle relative to the central ray will miss the converter.
The desirable total rate of photons on the converter is $10^{13}$ per second.
The photon flux results in an event rate of the $e^+e^-$ pairs of 10\,MHz (with acceptable 10\% background for the 10\,ns time window of the coordinate detector). 
One can use a fast calorimeter for detection of the total energy of the $e^+e^-$ pair which will be used to select the 131 and 1321 photon lines. 
After electronically made reduction of the trigger rate for the 131 line by a factor of 10, the total event rate will be 3\,MHz.
Here we assumed that data-acquisition electronics will be able to do on-line accumulation of the photon spectra at a very large rate because of a simple algorithm of the data analysis.
The statistics-accumulation will be 1.5\,MHz in each line.
With an estimated relative energy resolution for each photon of $10^{-4}$, the accuracy for the average photon energy 
$10^{-8}$ will be provided in one hundred seconds and similar precision will be achieved for the beam energy, see Equation~\ref{eq:131}. This corresponds to high sensitivity to one-way MAS variations.
Over 24 hours, the limit on $\delta c1_{ph}/c1_{ph}$ will be better than $10^{-16}$ (per Equation~\ref{eq:c1}), and  with a one- to two-month run it is likely to 
reach $\delta c1_{ph/}c1_{ph}$ on the level of $1 \cdot 10^{-17}$. 
We note that, since the converter (see Fig.~\ref{fig:PS}) interacts only with a small (ppm-level) fraction of the photon beam, the setup is well-suited for parasitic operation.

\section{Two photon beams in the opposite directions}
\label{sec:configuration_B}
Due to the beam energy instability, the LLI violation effect 
can not be easily observed with just one photon energy line, when the ion beam is moving in the clockwise direction (for example).
At the same time, thanks to the second beam of LHC, which is moving in the counter clockwise direction, the complementary GF setup will provide a photon beam moving in the direction Left.
The LLI violation effect (per SME) for the Left photon beam has an opposite sign from that for the Right photon beam. 
So, accordingly the photon energies ratio, $E_{131}^R/E_{131}^L$, will reveal the anisotropy of the photon one-way MAS.
The values of beam energies of two LHC beams could be slightly different but the relative variations are similar and the energy proposed below cross calibration is applicable (see section \ref{sec:energy}).

\section{The LLI test of the hadron maximum attainable speed}
\label{sec:configuration_C}

As it has been shown above, photon-energy measurement allows accurate determination of the Lorentz factor for the photon energy transformation between the lab frame and the ion (beam) rest-frame.
The Lorentz factor of such transformation, $1/\sqrt{1-(v_{beam}/c1_{ph})^2}$, will be measured to the level of $10^{-8}$ precision every one hundred seconds.
The value of the sidereal time variation of the photon MAS ($\delta c1_{ph}$) will be obtained in the proposed experiment with a precision of $10^{-16}$ or better, which means that, after correction using the measured photon-beam energy, the beam speed $v_{beam}$ will be effectively stable on a comparable level.

The LHC storage ring is equipped for high-precision measurements of the beam momentum, and especially its relative variation~\cite{Evans_2008}, via beam deflection in the stable bending magnets with the beam-position monitors.
The relative accuracy for the beam momentum is $10^{-6}$ with 10\,s seconds of data taking~\cite{JW2021aa}.
The stable beam speed allows one to study sidereal time variation of the maximum attainable speed for an ion, the hadron MAS, via comparison of the observed beam momentum from deflection of the beam trajectory in the ring magnets and the expected momentum for the  Lorentz factor   
$1/\sqrt{1-(v_{beam}/c1_h)^2}$.
The precision for the sidereal time variation of $c1_h$ will be at the $10^{-14}$ level or better.
  This configuration of the experiment could be a part of the measurement presented in Section~\ref{sec:configuration_A}.

\section{Conclusion}

The proposed Doppler experiment at the Gamma Factory will allow one to reach a sensitivity to variation of the photon maximum attainable speed
with sidereal time at the level of $10^{-17}$, which is orders of magnitude more advanced than the current best laboratory-based experiments. 
These projections are summarized in Table~\ref{tab:summary}.
\begin{table}[ht]
\addtolength{\tabcolsep}{+3pt}
\renewcommand{\arraystretch}{1.5}
\caption{Projected sensitivity of three proposed configurations in different analysis frameworks for one-way MAS of light and a hadron.}
\label{tab:summary}
\centering
    \begin{tabular}{|c|c|c|c|}
\hline  
Setup& $\delta c1/c$ of &  SME \cite{AKostel2011ab}&ARM \cite{Sonego2009ab} \\
\hline
A, Section~\ref{sec:configuration_A}   &   photon &    $10^{-17}$  &        $10^{-17}$       \\ 
B, Section~\ref{sec:configuration_B}   &   photon &    $10^{-17}$  &        $10^{-17}$       \\
C, Section~\ref{sec:configuration_C}   &   hadron &      -         &        $10^{-14}$       \\
\hline  
    \end{tabular}
\end{table} 
Variations of the hadron MAS can be searched for at a level of $10^{-14}$ or better using simultaneous measurement of the beam momentum in the same part of the ring orbit, which will also yield a unique result for the hadron one-way MAS.

\medskip
\textbf{Acknowledgements} \par 
The authors acknowledge helpful discussions with A.~Surzhykov and J.~Wenninger. 
We thank Mainz Institute of Theoretical Physics (MITP) and the participants of the MITP GF-2020 Workshop during which the ideas presented in this paper were developed. 
DB acknowledges support by the Cluster of Excellence PRISMA+ funded by the German Research Foundation (DFG) within the German Excellence Strategy (Project ID 39083149), by the European Research Council (ERC) under the European Union Horizon 2020 research and innovation program (project Dark-OST, grant agreement No 695405), and by the DFG Reinhart Koselleck project.

\bibliographystyle{apsrev4-2}
\bibliography{LLI-GF_arXiv.bib}

\begin{thebibliography}{34}%
\makeatletter
\providecommand \@ifxundefined [1]{%
 \@ifx{#1\undefined}
}%
\providecommand \@ifnum [1]{%
 \ifnum #1\expandafter \@firstoftwo
 \else \expandafter \@secondoftwo
 \fi
}%
\providecommand \@ifx [1]{%
 \ifx #1\expandafter \@firstoftwo
 \else \expandafter \@secondoftwo
 \fi
}%
\providecommand \natexlab [1]{#1}%
\providecommand \enquote  [1]{``#1''}%
\providecommand \bibnamefont  [1]{#1}%
\providecommand \bibfnamefont [1]{#1}%
\providecommand \citenamefont [1]{#1}%
\providecommand \href@noop [0]{\@secondoftwo}%
\providecommand \href [0]{\begingroup \@sanitize@url \@href}%
\providecommand \@href[1]{\@@startlink{#1}\@@href}%
\providecommand \@@href[1]{\endgroup#1\@@endlink}%
\providecommand \@sanitize@url [0]{\catcode `\\12\catcode `\$12\catcode
  `\&12\catcode `\#12\catcode `\^12\catcode `\_12\catcode `\%12\relax}%
\providecommand \@@startlink[1]{}%
\providecommand \@@endlink[0]{}%
\providecommand \url  [0]{\begingroup\@sanitize@url \@url }%
\providecommand \@url [1]{\endgroup\@href {#1}{\urlprefix }}%
\providecommand \urlprefix  [0]{URL }%
\providecommand \Eprint [0]{\href }%
\providecommand \doibase [0]{https://doi.org/}%
\providecommand \selectlanguage [0]{\@gobble}%
\providecommand \bibinfo  [0]{\@secondoftwo}%
\providecommand \bibfield  [0]{\@secondoftwo}%
\providecommand \translation [1]{[#1]}%
\providecommand \BibitemOpen [0]{}%
\providecommand \bibitemStop [0]{}%
\providecommand \bibitemNoStop [0]{.\EOS\space}%
\providecommand \EOS [0]{\spacefactor3000\relax}%
\providecommand \BibitemShut  [1]{\csname bibitem#1\endcsname}%
\let\auto@bib@innerbib\@empty
\bibitem [{\citenamefont {Mattingly}(2005)}]{Mattinly2005ab}%
  \BibitemOpen
  \bibfield  {author} {\bibinfo {author} {\bibfnamefont {D.}~\bibnamefont
  {Mattingly}},\ }\href {https://doi.org/10.12942/lrr-2005-5} {\bibfield
  {journal} {\bibinfo  {journal} {Living Reviews in Relativity}\ }\textbf
  {\bibinfo {volume} {8}},\ \bibinfo {pages} {5} (\bibinfo {year}
  {2005})}\BibitemShut {NoStop}%
\bibitem [{\citenamefont {Kosteleck\'y}\ and\ \citenamefont
  {Russell}(2011)}]{AKostel2011ab}%
  \BibitemOpen
  \bibfield  {author} {\bibinfo {author} {\bibfnamefont {V.~A.}\ \bibnamefont
  {Kosteleck\'y}}\ and\ \bibinfo {author} {\bibfnamefont {N.}~\bibnamefont
  {Russell}},\ }\href {https://doi.org/10.1103/RevModPhys.83.11} {\bibfield
  {journal} {\bibinfo  {journal} {Rev. Mod. Phys.}\ }\textbf {\bibinfo {volume}
  {83}},\ \bibinfo {pages} {11} (\bibinfo {year} {2011})}\BibitemShut {NoStop}%
\bibitem [{\citenamefont {Will}(2014)}]{Will2014ab}%
  \BibitemOpen
  \bibfield  {author} {\bibinfo {author} {\bibfnamefont {C.}~\bibnamefont
  {Will}},\ }\href {https://doi.org/10.12942/lrr-2014-4} {\bibfield  {journal}
  {\bibinfo  {journal} {Living Reviews in Relativity}\ }\textbf {\bibinfo
  {volume} {17}},\ \bibinfo {pages} {4} (\bibinfo {year} {2014})}\BibitemShut
  {NoStop}%
\bibitem [{\citenamefont {Robertson}(1949)}]{HRobert1949ab}%
  \BibitemOpen
  \bibfield  {author} {\bibinfo {author} {\bibfnamefont {H.~P.}\ \bibnamefont
  {Robertson}},\ }\href {https://doi.org/10.1103/RevModPhys.21.378} {\bibfield
  {journal} {\bibinfo  {journal} {Rev. Mod. Phys.}\ }\textbf {\bibinfo {volume}
  {21}},\ \bibinfo {pages} {378} (\bibinfo {year} {1949})}\BibitemShut
  {NoStop}%
\bibitem [{\citenamefont {Monsouri}\ and\ \citenamefont
  {Sexl}(1977)}]{Mansouri1977ab}%
  \BibitemOpen
  \bibfield  {author} {\bibinfo {author} {\bibfnamefont {R.}~\bibnamefont
  {Monsouri}}\ and\ \bibinfo {author} {\bibfnamefont {R.}~\bibnamefont
  {Sexl}},\ }\href {https://doi.org/10.1007/BF00762634} {\bibfield  {journal}
  {\bibinfo  {journal} {General Relativity and Gravitation}\ }\textbf {\bibinfo
  {volume} {8}},\ \bibinfo {pages} {497, 515, 809} (\bibinfo {year}
  {1977})}\BibitemShut {NoStop}%
\bibitem [{\citenamefont {Colladay}\ and\ \citenamefont
  {Kosteleck\'y}(1997)}]{Colladay1997a}%
  \BibitemOpen
  \bibfield  {author} {\bibinfo {author} {\bibfnamefont {D.}~\bibnamefont
  {Colladay}}\ and\ \bibinfo {author} {\bibfnamefont {V.~A.}\ \bibnamefont
  {Kosteleck\'y}},\ }\href {https://doi.org/10.1103/PhysRevD.55.6760}
  {\bibfield  {journal} {\bibinfo  {journal} {Phys. Rev. D}\ }\textbf {\bibinfo
  {volume} {55}},\ \bibinfo {pages} {6760} (\bibinfo {year}
  {1997})}\BibitemShut {NoStop}%
\bibitem [{\citenamefont {Michelson}\ and\ \citenamefont
  {Morley}(1887)}]{MM1887ab}%
  \BibitemOpen
  \bibfield  {author} {\bibinfo {author} {\bibfnamefont {A.}~\bibnamefont
  {Michelson}}\ and\ \bibinfo {author} {\bibfnamefont {E.}~\bibnamefont
  {Morley}},\ }\href {https://doi.org/10.2475/ajs.s3-34.203.333} {\bibfield
  {journal} {\bibinfo  {journal} {American Journal of Science}\ }\textbf
  {\bibinfo {volume} {34}},\ \bibinfo {pages} {333} (\bibinfo {year}
  {1887})}\BibitemShut {NoStop}%
\bibitem [{\citenamefont {Colladay}\ and\ \citenamefont
  {Kosteleck\'y}(1998)}]{Colladay1998b}%
  \BibitemOpen
  \bibfield  {author} {\bibinfo {author} {\bibfnamefont {D.}~\bibnamefont
  {Colladay}}\ and\ \bibinfo {author} {\bibfnamefont {V.~A.}\ \bibnamefont
  {Kosteleck\'y}},\ }\href {https://doi.org/10.1103/PhysRevD.58.116002}
  {\bibfield  {journal} {\bibinfo  {journal} {Phys. Rev. D}\ }\textbf {\bibinfo
  {volume} {58}},\ \bibinfo {pages} {116002} (\bibinfo {year}
  {1998})}\BibitemShut {NoStop}%
\bibitem [{\citenamefont {Coleman}\ and\ \citenamefont
  {Glashow}(1999)}]{Col1999ab}%
  \BibitemOpen
  \bibfield  {author} {\bibinfo {author} {\bibfnamefont {S.}~\bibnamefont
  {Coleman}}\ and\ \bibinfo {author} {\bibfnamefont {S.~L.}\ \bibnamefont
  {Glashow}},\ }\href {https://doi.org/10.1103/PhysRevD.59.116008} {\bibfield
  {journal} {\bibinfo  {journal} {Phys. Rev. D}\ }\textbf {\bibinfo {volume}
  {59}},\ \bibinfo {pages} {116008} (\bibinfo {year} {1999})}\BibitemShut
  {NoStop}%
\bibitem [{\citenamefont {Einstein}(1905)}]{AE1905ab}%
  \BibitemOpen
  \bibfield  {author} {\bibinfo {author} {\bibfnamefont {A.}~\bibnamefont
  {Einstein}},\ }\href {https://doi.org/10.1002/andp.200590006} {\bibfield
  {journal} {\bibinfo  {journal} {Annalen der Physik}\ }\textbf {\bibinfo
  {volume} {17}},\ \bibinfo {pages} {891} (\bibinfo {year} {1905})}\BibitemShut
  {NoStop}%
\bibitem [{\citenamefont {Ives}\ and\ \citenamefont
  {Stilwell}(1938)}]{Ives1938ab}%
  \BibitemOpen
  \bibfield  {author} {\bibinfo {author} {\bibfnamefont {H.}~\bibnamefont
  {Ives}}\ and\ \bibinfo {author} {\bibfnamefont {G.}~\bibnamefont
  {Stilwell}},\ }\href {https://doi.org/10.1364/JOSA.28.000215} {\bibfield
  {journal} {\bibinfo  {journal} {Journal of the Optical Society of America}\
  }\textbf {\bibinfo {volume} {28}},\ \bibinfo {pages} {215} (\bibinfo {year}
  {1938})}\BibitemShut {NoStop}%
\bibitem [{\citenamefont {Botermann}\ and\ \citenamefont
  {other}(2014)}]{Botermann2014ab}%
  \BibitemOpen
  \bibfield  {author} {\bibinfo {author} {\bibfnamefont {B.}~\bibnamefont
  {Botermann}}\ and\ \bibinfo {author} {\bibnamefont {other}},\ }\href
  {https://doi.org/10.1103/PhysRevLett.113.120405} {\bibfield  {journal}
  {\bibinfo  {journal} {Phys. Rev. Lett.}\ }\textbf {\bibinfo {volume} {113}},\
  \bibinfo {pages} {120405} (\bibinfo {year} {2014})}\BibitemShut {NoStop}%
\bibitem [{\citenamefont {Gurzadyan}\ and\ \citenamefont
  {Margarian}(1996)}]{Gurzadyan1996ab}%
  \BibitemOpen
  \bibfield  {author} {\bibinfo {author} {\bibfnamefont {V.}~\bibnamefont
  {Gurzadyan}}\ and\ \bibinfo {author} {\bibfnamefont {A.}~\bibnamefont
  {Margarian}},\ }\href {https://doi.org/10.1088/0031-8949/53/5/001} {\bibfield
   {journal} {\bibinfo  {journal} {Physica Scripta}\ }\textbf {\bibinfo
  {volume} {53}},\ \bibinfo {pages} {513} (\bibinfo {year} {1996})}\BibitemShut
  {NoStop}%
\bibitem [{\citenamefont {Bocquet}\ \emph {et~al.}(2010)\citenamefont {Bocquet}
  \emph {et~al.}}]{Bocquet:2010ke}%
  \BibitemOpen
  \bibfield  {author} {\bibinfo {author} {\bibfnamefont {J.-P.}\ \bibnamefont
  {Bocquet}} \emph {et~al.},\ }\href
  {https://doi.org/10.1103/PhysRevLett.104.241601} {\bibfield  {journal}
  {\bibinfo  {journal} {Phys. Rev. Lett.}\ }\textbf {\bibinfo {volume} {104}},\
  \bibinfo {pages} {241601} (\bibinfo {year} {2010})}\BibitemShut {NoStop}%
\bibitem [{\citenamefont {Wojtsekhowski}(2014)}]{Wojt2014ab}%
  \BibitemOpen
  \bibfield  {author} {\bibinfo {author} {\bibfnamefont {B.}~\bibnamefont
  {Wojtsekhowski}},\ }\href {https://doi.org/10.1209/0295-5075/108/31001}
  {\bibfield  {journal} {\bibinfo  {journal} {{EPL} (Europhysics Letters)}\
  }\textbf {\bibinfo {volume} {108}},\ \bibinfo {pages} {31001} (\bibinfo
  {year} {2014})}\BibitemShut {NoStop}%
\bibitem [{\citenamefont {Bergan}\ \emph {et~al.}(2020)\citenamefont {Bergan}
  \emph {et~al.}}]{Bergan:2019lff}%
  \BibitemOpen
  \bibfield  {author} {\bibinfo {author} {\bibfnamefont {W.}~\bibnamefont
  {Bergan}} \emph {et~al.},\ }\href
  {https://doi.org/10.1103/PhysRevD.101.032004} {\bibfield  {journal} {\bibinfo
   {journal} {Phys. Rev. D}\ }\textbf {\bibinfo {volume} {101}},\ \bibinfo
  {pages} {032004} (\bibinfo {year} {2020})},\ \Eprint
  {https://arxiv.org/abs/2001.00999} {arXiv:2001.00999 [hep-ex]} \BibitemShut
  {NoStop}%
\bibitem [{\citenamefont {Sonego}\ and\ \citenamefont
  {Pin}(2009)}]{Sonego2009ab}%
  \BibitemOpen
  \bibfield  {author} {\bibinfo {author} {\bibfnamefont {S.}~\bibnamefont
  {Sonego}}\ and\ \bibinfo {author} {\bibfnamefont {M.}~\bibnamefont {Pin}},\
  }\href {https://doi.org/10.1063/1.3104065} {\bibfield  {journal} {\bibinfo
  {journal} {Journal of Mathematical Physics}\ }\textbf {\bibinfo {volume}
  {50}},\ \bibinfo {pages} {042902} (\bibinfo {year} {2009})}\BibitemShut
  {NoStop}%
\bibitem [{\citenamefont {Krasny}(2015)}]{krasny2015gamma}%
  \BibitemOpen
  \bibfield  {author} {\bibinfo {author} {\bibfnamefont {M.~W.}\ \bibnamefont
  {Krasny}},\ }\href {https://arxiv.org/abs/1511.07794} {\bibfield  {journal}
  {\bibinfo  {journal} {arXiv:1511.07794}\ } (\bibinfo {year}
  {2015})}\BibitemShut {NoStop}%
\bibitem [{\citenamefont {Krasny}\ \emph {et~al.}(2018)\citenamefont {Krasny},
  \citenamefont {Antsiferov}, \citenamefont {Apyan}, \citenamefont
  {Alikhanyan}, \citenamefont {Bessonov}, \citenamefont {Shevelko},
  \citenamefont {Lebedev}, \citenamefont {Budker}, \citenamefont {Cassou},
  \citenamefont {Chaikovska}, \citenamefont {Chehab}, \citenamefont {Dupraz},
  \citenamefont {Martens}, \citenamefont {Zomer}, \citenamefont {Castelli},
  \citenamefont {Curatolo}, \citenamefont {Petrillo}, \citenamefont {Serafini},
  \citenamefont {Bartosik}, \citenamefont {Biancacci}, \citenamefont
  {Czodrowski}, \citenamefont {Goddard}, \citenamefont {Jowett}, \citenamefont
  {Alemany~Fernandez}, \citenamefont {Hirlander}, \citenamefont {Kersevan},
  \citenamefont {Kowalska}, \citenamefont {Lamont}, \citenamefont {Manglunki},
  \citenamefont {Petrenko}, \citenamefont {Schaumann}, \citenamefont
  {Yin-Vallgren}, \citenamefont {Zimmermann}, \citenamefont {Bieron},
  \citenamefont {Dzierzega}, \citenamefont {Placzek}, \citenamefont {Pustelny},
  \citenamefont {Kroeger}, \citenamefont {Stoehlker}, \citenamefont {Weber},
  \citenamefont {Wu},\ and\ \citenamefont {Zolotorev}}]{Krasny2018}%
  \BibitemOpen
  \bibfield  {author} {\bibinfo {author} {\bibfnamefont {M.~W.}\ \bibnamefont
  {Krasny}}, \bibinfo {author} {\bibfnamefont {P.~S.}\ \bibnamefont
  {Antsiferov}}, \bibinfo {author} {\bibfnamefont {A.}~\bibnamefont {Apyan}},
  \bibinfo {author} {\bibfnamefont {A.~I.}\ \bibnamefont {Alikhanyan}},
  \bibinfo {author} {\bibfnamefont {E.~G.}\ \bibnamefont {Bessonov}}, \bibinfo
  {author} {\bibfnamefont {V.~P.}\ \bibnamefont {Shevelko}}, \bibinfo {author}
  {\bibfnamefont {P.~N.}\ \bibnamefont {Lebedev}}, \bibinfo {author}
  {\bibfnamefont {D.}~\bibnamefont {Budker}}, \bibinfo {author} {\bibfnamefont
  {K.}~\bibnamefont {Cassou}}, \bibinfo {author} {\bibfnamefont
  {I.}~\bibnamefont {Chaikovska}}, \bibinfo {author} {\bibfnamefont
  {R.}~\bibnamefont {Chehab}}, \bibinfo {author} {\bibfnamefont
  {K.}~\bibnamefont {Dupraz}}, \bibinfo {author} {\bibfnamefont
  {A.}~\bibnamefont {Martens}}, \bibinfo {author} {\bibfnamefont
  {F.}~\bibnamefont {Zomer}}, \bibinfo {author} {\bibfnamefont
  {F.}~\bibnamefont {Castelli}}, \bibinfo {author} {\bibfnamefont
  {C.}~\bibnamefont {Curatolo}}, \bibinfo {author} {\bibfnamefont
  {V.}~\bibnamefont {Petrillo}}, \bibinfo {author} {\bibfnamefont
  {L.}~\bibnamefont {Serafini}}, \bibinfo {author} {\bibfnamefont
  {H.}~\bibnamefont {Bartosik}}, \bibinfo {author} {\bibfnamefont
  {N.}~\bibnamefont {Biancacci}}, \bibinfo {author} {\bibfnamefont
  {P.}~\bibnamefont {Czodrowski}}, \bibinfo {author} {\bibfnamefont
  {B.}~\bibnamefont {Goddard}}, \bibinfo {author} {\bibfnamefont {J.~M.}\
  \bibnamefont {Jowett}}, \bibinfo {author} {\bibfnamefont {R.}~\bibnamefont
  {Alemany~Fernandez}}, \bibinfo {author} {\bibfnamefont {S.}~\bibnamefont
  {Hirlander}}, \bibinfo {author} {\bibfnamefont {R.}~\bibnamefont {Kersevan}},
  \bibinfo {author} {\bibfnamefont {M.}~\bibnamefont {Kowalska}}, \bibinfo
  {author} {\bibfnamefont {M.}~\bibnamefont {Lamont}}, \bibinfo {author}
  {\bibfnamefont {D.}~\bibnamefont {Manglunki}}, \bibinfo {author}
  {\bibfnamefont {A.}~\bibnamefont {Petrenko}}, \bibinfo {author}
  {\bibfnamefont {M.}~\bibnamefont {Schaumann}}, \bibinfo {author}
  {\bibfnamefont {C.}~\bibnamefont {Yin-Vallgren}}, \bibinfo {author}
  {\bibfnamefont {F.}~\bibnamefont {Zimmermann}}, \bibinfo {author}
  {\bibfnamefont {J.}~\bibnamefont {Bieron}}, \bibinfo {author} {\bibfnamefont
  {K.}~\bibnamefont {Dzierzega}}, \bibinfo {author} {\bibfnamefont
  {W.}~\bibnamefont {Placzek}}, \bibinfo {author} {\bibfnamefont
  {S.}~\bibnamefont {Pustelny}}, \bibinfo {author} {\bibfnamefont
  {F.}~\bibnamefont {Kroeger}}, \bibinfo {author} {\bibfnamefont
  {T.}~\bibnamefont {Stoehlker}}, \bibinfo {author} {\bibfnamefont
  {G.}~\bibnamefont {Weber}}, \bibinfo {author} {\bibfnamefont {Y.~K.}\
  \bibnamefont {Wu}},\ and\ \bibinfo {author} {\bibfnamefont {M.~S.}\
  \bibnamefont {Zolotorev}},\ }in\ \href
  {https://doi.org/doi:10.18429/JACoW-IPAC2018-WEYGBD3} {\emph {\bibinfo
  {booktitle} {Proc. 9th International Particle Accelerator Conference
  (IPAC'18), Vancouver, BC, Canada, April 29-May 4, 2018}}},\ \bibinfo {series
  and number} {\bibinfo {series} {International Particle Accelerator
  Conference}\ No.~\bibinfo {number} {9}}\ (\bibinfo  {publisher} {Joint
  Accelerator Conferences Website},\ \bibinfo {year} {2018})\ pp.\ \bibinfo
  {pages} {1780--1783}\BibitemShut {NoStop}%
\bibitem [{\citenamefont {Ispirian}\ and\ \citenamefont
  {Margarian}(1973)}]{ISPIRIAN1973377}%
  \BibitemOpen
  \bibfield  {author} {\bibinfo {author} {\bibfnamefont {K.}~\bibnamefont
  {Ispirian}}\ and\ \bibinfo {author} {\bibfnamefont {A.}~\bibnamefont
  {Margarian}},\ }\href
  {https://doi.org/https://doi.org/10.1016/0375-9601(73)90791-3} {\bibfield
  {journal} {\bibinfo  {journal} {Physics Letters A}\ }\textbf {\bibinfo
  {volume} {44}},\ \bibinfo {pages} {377 } (\bibinfo {year}
  {1973})}\BibitemShut {NoStop}%
\bibitem [{\citenamefont {Budker}\ \emph {et~al.}(2020)\citenamefont {Budker},
  \citenamefont {Crespo López-Urrutia}, \citenamefont {Derevianko},
  \citenamefont {Flambaum}, \citenamefont {Krasny}, \citenamefont {Petrenko},
  \citenamefont {Pustelny}, \citenamefont {Surzhykov}, \citenamefont
  {Yerokhin},\ and\ \citenamefont {Zolotorev}}]{budker2020atomic}%
  \BibitemOpen
  \bibfield  {author} {\bibinfo {author} {\bibfnamefont {D.}~\bibnamefont
  {Budker}}, \bibinfo {author} {\bibfnamefont {J.~R.}\ \bibnamefont {Crespo
  López-Urrutia}}, \bibinfo {author} {\bibfnamefont {A.}~\bibnamefont
  {Derevianko}}, \bibinfo {author} {\bibfnamefont {V.~V.}\ \bibnamefont
  {Flambaum}}, \bibinfo {author} {\bibfnamefont {M.~W.}\ \bibnamefont
  {Krasny}}, \bibinfo {author} {\bibfnamefont {A.}~\bibnamefont {Petrenko}},
  \bibinfo {author} {\bibfnamefont {S.}~\bibnamefont {Pustelny}}, \bibinfo
  {author} {\bibfnamefont {A.}~\bibnamefont {Surzhykov}}, \bibinfo {author}
  {\bibfnamefont {V.~A.}\ \bibnamefont {Yerokhin}},\ and\ \bibinfo {author}
  {\bibfnamefont {M.}~\bibnamefont {Zolotorev}},\ }\href
  {https://doi.org/https://doi.org/10.1002/andp.202000204} {\bibfield
  {journal} {\bibinfo  {journal} {Annalen der Physik}\ }\textbf {\bibinfo
  {volume} {532}},\ \bibinfo {pages} {2000204} (\bibinfo {year} {2020})},\
  \Eprint
  {https://arxiv.org/abs/https://onlinelibrary.wiley.com/doi/pdf/10.1002/andp.202000204}
  {https://onlinelibrary.wiley.com/doi/pdf/10.1002/andp.202000204} \BibitemShut
  {NoStop}%
\bibitem [{\citenamefont {Jentschura}\ and\ \citenamefont
  {Surzhykov}(2008)}]{PhysRevA.77.042507}%
  \BibitemOpen
  \bibfield  {author} {\bibinfo {author} {\bibfnamefont {U.~D.}\ \bibnamefont
  {Jentschura}}\ and\ \bibinfo {author} {\bibfnamefont {A.}~\bibnamefont
  {Surzhykov}},\ }\href {https://doi.org/10.1103/PhysRevA.77.042507} {\bibfield
   {journal} {\bibinfo  {journal} {Phys. Rev. A}\ }\textbf {\bibinfo {volume}
  {77}},\ \bibinfo {pages} {042507} (\bibinfo {year} {2008})}\BibitemShut
  {NoStop}%
\bibitem [{\citenamefont {Surzhykov}\ \emph {et~al.}(2005)\citenamefont
  {Surzhykov}, \citenamefont {Koval},\ and\ \citenamefont
  {Fritzsche}}]{SURZHYKOV2005139}%
  \BibitemOpen
  \bibfield  {author} {\bibinfo {author} {\bibfnamefont {A.}~\bibnamefont
  {Surzhykov}}, \bibinfo {author} {\bibfnamefont {P.}~\bibnamefont {Koval}},\
  and\ \bibinfo {author} {\bibfnamefont {S.}~\bibnamefont {Fritzsche}},\ }\href
  {https://doi.org/https://doi.org/10.1016/j.cpc.2004.09.004} {\bibfield
  {journal} {\bibinfo  {journal} {Computer Physics Communications}\ }\textbf
  {\bibinfo {volume} {165}},\ \bibinfo {pages} {139} (\bibinfo {year}
  {2005})}\BibitemShut {NoStop}%
\bibitem [{\citenamefont {McConnell}\ \emph {et~al.}(2010)\citenamefont
  {McConnell}, \citenamefont {Fritzsche},\ and\ \citenamefont
  {Surzhykov}}]{MCCONNELL2010711}%
  \BibitemOpen
  \bibfield  {author} {\bibinfo {author} {\bibfnamefont {S.}~\bibnamefont
  {McConnell}}, \bibinfo {author} {\bibfnamefont {S.}~\bibnamefont
  {Fritzsche}},\ and\ \bibinfo {author} {\bibfnamefont {A.}~\bibnamefont
  {Surzhykov}},\ }\href
  {https://doi.org/https://doi.org/10.1016/j.cpc.2009.11.010} {\bibfield
  {journal} {\bibinfo  {journal} {Computer Physics Communications}\ }\textbf
  {\bibinfo {volume} {181}},\ \bibinfo {pages} {711} (\bibinfo {year}
  {2010})}\BibitemShut {NoStop}%
\bibitem [{\citenamefont {Popov}\ and\ \citenamefont
  {Maiorova}(2017)}]{popov2017relativistic}%
  \BibitemOpen
  \bibfield  {author} {\bibinfo {author} {\bibfnamefont {R.}~\bibnamefont
  {Popov}}\ and\ \bibinfo {author} {\bibfnamefont {A.}~\bibnamefont
  {Maiorova}},\ }\href {https://doi.org/10.1134/S0030400X1703016X} {\bibfield
  {journal} {\bibinfo  {journal} {Optics and Spectroscopy}\ }\textbf {\bibinfo
  {volume} {122}},\ \bibinfo {pages} {366} (\bibinfo {year}
  {2017})}\BibitemShut {NoStop}%
\bibitem [{\citenamefont {Surzhykov}(2021)}]{andreysur}%
  \BibitemOpen
  \bibfield  {author} {\bibinfo {author} {\bibfnamefont {A.}~\bibnamefont
  {Surzhykov}}} (\bibinfo {year} {2021})\BibitemShut {NoStop}%
\bibitem [{\citenamefont {Arnaudon}\ \emph {et~al.}(1995)\citenamefont
  {Arnaudon} \emph {et~al.}}]{ARNAUDON1995249}%
  \BibitemOpen
  \bibfield  {author} {\bibinfo {author} {\bibfnamefont {L.}~\bibnamefont
  {Arnaudon}} \emph {et~al.},\ }\href
  {https://doi.org/https://doi.org/10.1016/0168-9002(94)01526-0} {\bibfield
  {journal} {\bibinfo  {journal} {Nuclear Instruments and Methods in Physics
  Research Section A}\ }\textbf {\bibinfo {volume} {357}},\ \bibinfo {pages}
  {249 } (\bibinfo {year} {1995})}\BibitemShut {NoStop}%
\bibitem [{\citenamefont {Nagel}\ \emph {et~al.}(2015)\citenamefont {Nagel}
  \emph {et~al.}}]{Nagel_2015}%
  \BibitemOpen
  \bibfield  {author} {\bibinfo {author} {\bibfnamefont {M.}~\bibnamefont
  {Nagel}} \emph {et~al.},\ }\href {https://doi.org/10.1038/ncomms9174}
  {\bibfield  {journal} {\bibinfo  {journal} {Nature Communication}\ }\textbf
  {\bibinfo {volume} {6}},\ \bibinfo {pages} {8174} (\bibinfo {year}
  {2015})}\BibitemShut {NoStop}%
\bibitem [{\citenamefont {Wojtsekhowski}(2020)}]{BW2020ab}%
  \BibitemOpen
  \bibfield  {author} {\bibinfo {author} {\bibfnamefont {B.}~\bibnamefont
  {Wojtsekhowski}}} (\bibinfo {year} {2020}),\ \bibinfo {note} {report at the
  workshop "Physics Opportunities with the Gamma Factory"}\BibitemShut
  {NoStop}%
\bibitem [{\citenamefont {Golubnichiy}\ \emph {et~al.}(1969)\citenamefont
  {Golubnichiy} \emph {et~al.}}]{Golubnichiy:1969ui}%
  \BibitemOpen
  \bibfield  {author} {\bibinfo {author} {\bibfnamefont {P.~I.}\ \bibnamefont
  {Golubnichiy}} \emph {et~al.},\ }\href
  {https://doi.org/https://doi.org/10.1016/0029-554X(69)90536-9} {\bibfield
  {journal} {\bibinfo  {journal} {Nuclear Instruments and Methods}\ }\textbf
  {\bibinfo {volume} {67}},\ \bibinfo {pages} {22} (\bibinfo {year}
  {1969})}\BibitemShut {NoStop}%
\bibitem [{\citenamefont {Battaglieri}\ \emph {et~al.}(2015)\citenamefont
  {Battaglieri} \emph {et~al.}}]{BATTAGLIERI201591}%
  \BibitemOpen
  \bibfield  {author} {\bibinfo {author} {\bibfnamefont {M.}~\bibnamefont
  {Battaglieri}} \emph {et~al.},\ }\href
  {https://doi.org/https://doi.org/10.1016/j.nima.2014.12.017} {\bibfield
  {journal} {\bibinfo  {journal} {Nuclear Instruments and Methods in Physics
  Research Section A}\ }\textbf {\bibinfo {volume} {777}},\ \bibinfo {pages}
  {91} (\bibinfo {year} {2015})}\BibitemShut {NoStop}%
\bibitem [{\citenamefont {Dewey}\ \emph {et~al.}(2006)\citenamefont {Dewey}
  \emph {et~al.}}]{PhysRevC.73.044303}%
  \BibitemOpen
  \bibfield  {author} {\bibinfo {author} {\bibfnamefont {M.~S.}\ \bibnamefont
  {Dewey}} \emph {et~al.},\ }\href {https://doi.org/10.1103/PhysRevC.73.044303}
  {\bibfield  {journal} {\bibinfo  {journal} {Phys. Rev. C}\ }\textbf {\bibinfo
  {volume} {73}},\ \bibinfo {pages} {044303} (\bibinfo {year}
  {2006})}\BibitemShut {NoStop}%
\bibitem [{\citenamefont {Evans}\ and\ \citenamefont
  {Bryant}(2008)}]{Evans_2008}%
  \BibitemOpen
  \bibfield  {author} {\bibinfo {author} {\bibfnamefont {L.}~\bibnamefont
  {Evans}}\ and\ \bibinfo {author} {\bibfnamefont {P.}~\bibnamefont {Bryant}},\
  }\href {https://doi.org/10.1088/1748-0221/3/08/s08001} {\bibfield  {journal}
  {\bibinfo  {journal} {Journal of Instrumentation}\ }\textbf {\bibinfo
  {volume} {3}}\bibinfo  {number} { (08)},\ \bibinfo {pages}
  {S08001}}\BibitemShut {NoStop}%
\bibitem [{\citenamefont {Wenninger}(2021)}]{JW2021aa}%
  \BibitemOpen
\bibfield  {number} {  }\bibfield  {author} {\bibinfo {author} {\bibfnamefont
  {J.}~\bibnamefont {Wenninger}}} (\bibinfo {year} {2021})\BibitemShut
  {NoStop}%
\end{thebibliography}%

\end{document}